\documentclass[a4paper]{article}

\usepackage{amsfonts}
\usepackage{graphicx}
\usepackage{amsbsy}
\usepackage{amsmath}
\usepackage{theorem}
\usepackage{dcolumn}
\usepackage{enumerate}
\usepackage{subfigure}
\usepackage [sorting=none]{biblatex}
\bibliography{essay}
\usepackage[ansinew]{inputenc}
\begin{document}

\begin{center} John Paul Adrian Glaubitz \end{center}
\begin{center} glaubitz@physik.fu-berlin.de \end{center}
\begin{center}\LARGE \textbf{Modern consumerism and the waste problem} \end{center}
\begin{center}July, 15 2011\end{center}
\begin{center} \textbf{Department of Physics} \end{center} \vspace{-6mm}
\begin{center} \textbf{Faculty of Mathematics and Natural Sciences} \end{center}\vspace{-6mm}
\begin{center} \textbf{University of Oslo} \end{center} \vskip5mm


\section{Introduction}
\label{sec:introduction}
With the advance of industrial mass production, modern
micro-electronics and computers, the intervals between the release of
new generations of consumer products have been dramatically reduced and
so have their lifetime cycles.

While it was very natural in the post-war era, that sophisticated
consumer products like television sets and stereo equipment would not be
replaced with a new product until they break, and usually beyond that point
since it was very common to have a broken television set serviced, the
habits of consumers have changed during the last quarter of the 20th
century. The variety of television sets, radios, stereos, phones and
computers available on the market has surged and more and more vendors
are competing against each other to win the favor of consumers.

A modern consumer product, like Apple's famous iPhone
has a market life of approximately one year \cite{listofiphonemodels}
until a successor is announced and subsequently pushed onto the
market. Usually these new generations bring a bunch of new features,
have a higher performance (CPU speed, battery life, radio data rates)
while maintaining the price or becoming even cheaper, thus the
consumer greatly benefits from the reduced lifetime cycle of these
products.

On the other hand, electronic devices not only require  a lot of
of Earth's limited resources for their production (including rare
metals like tantalum and indium), but their production processes are a
major source for harmful climate gases like carbon dioxide and toxic
waste like heavy metal alloys, acids and alkalis. And last but not
least is every obsoleted iPhone a candidate for waste facilities
unless consumers are going to sell them on the second hand market.

While we can not expect consumers and manufacturers to go back to the
early days of consumer products where lifetime cycles reached up to
20 years, the world record being the famous ``Centennial Lightbulb''
in Livermore, California in the United States, which has been lit for
over 100 years \cite{centenniallight}, it is certainly about time to
rethink modern consumerism with regard to responsibility to future
generations.

In this essay I would like to discuss the problem using the example of
Apple Inc. and their famous iPhone and the manufacturers of inkjet
printers. I will examine possible ways to improve the
situation with the help of viable alternatives and ways to prolong
lifetime cycles as well as if and how much governments should try to
tackle the problems of planned obsolescence and waste production
by passing of stricter laws to regulate the markets.

\section{Proprietary standards and patents}
\label{sec:proprietary_standards}

When Apple introduced their first iPod, in 2001 and, 6 years later in
2007, the iPhone, they set to revolutionize the markets for portable
music players as well as smart phones. Thanks to their unique design
and ease of use, iPods quickly evolved as world's most popular
portable players, nowadays account for over 70\% of all players sold
world-wide \cite{ipod_market_share}. While not as dominant within its
market as the iPod is in the market for portable music players
(market share as of the first quarter of 2011: 16.8\%
\cite{iphone_market_share}), the iPhone is still very popular and has
had a strong impact on its market, setting a standard for most mobile
phones to be released later on.

\begin{figure}
\begin{center}
  \fbox{\includegraphics[width=.5\textwidth]{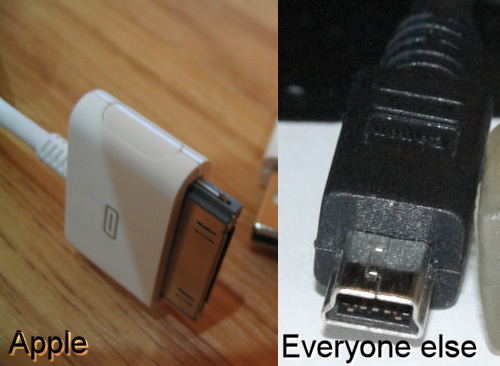}}
\end{center}
 \caption[Proprietary iPod connector]{Proprietary dock connector found
   in Apple's iPod and iPhone products (left) compared to a standard
   mini USB plug (right) found in most handheld devices by other
   manufacturers \cite{reddit_proprietary_ipod_cable}. Apple's
   proprietary connector makes it impossible to use cheap standard
   cables and chargers supplied with non-Apple brand handhelds.}
 \label{fig:ipod_proprietary_cable}
\end{figure}

With the introduction of the iPod and the iPhone, Apple introduced a
new, proprietary type of data and charge connector, the so-called
``30-pin dock connector'' \cite{ipod_dock_connector} which can
exclusively be found on Apple's handheld devices. While the design of
the 30-pin dock connector allows combining many connectors into a
single one and thus reducing the amount of connectors while still
remaining universal connectivity\footnote{Both the iPod and the iPhone
have, in fact, only a headphone jack and the 30-pin dock connector.},
the proprietary connector takes away the possibility to charge and
connect the iPod/iPhone with a standard USB cable as it can be done
with most other handheld devices on the market (see Figure
\ref{fig:ipod_proprietary_cable}).

Besides the practical difficulties it introduces, the use of such a
proprietary connector is also a potential source for avoidable
electronic waste, since all accessories and cables for the
iPod/iPhone cannot be used with non-Apple products if the consumer
decides to buy a different brand after the Apple handheld has reached
the end of its lifetime cycle.

Another very prominent example shown in the 2010 documentary movie
``The Light Bulb Conspiracy'' by Cosima Dannortizer
\cite{imdb_light_bulb_conspiracy} are consumer inkjet printers
produced by companies like HP, Canon and Lexmark. All these
manufacturers use proprietary types of ink cartridges which are
protected by international patents against reproduction by third-party
manufacturers \cite{canon_cartridge_patent, hp_cartridge_patent,
lexmark_cartridge_patent}, forcing the customer to buy the expensive
original cartridges.

\begin{figure}
\begin{center}
  \fbox{\includegraphics[width=.5\textwidth]{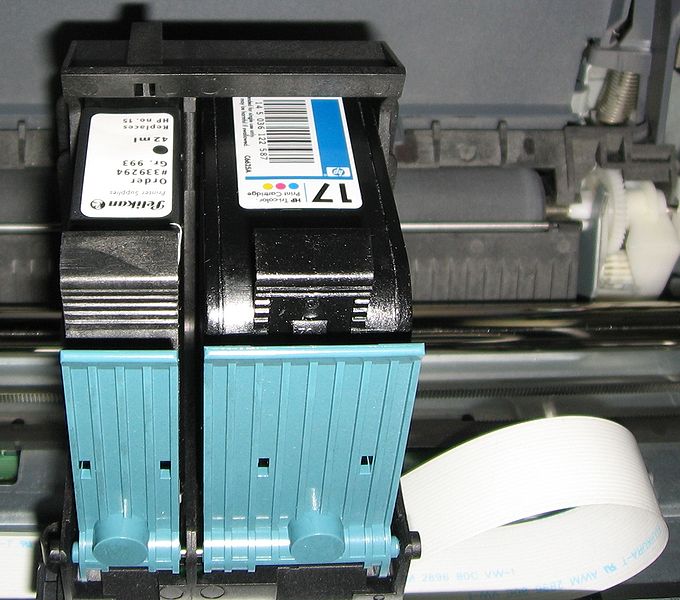}}
\end{center}
 \caption[Typical inkjet cartridge]{Typical inkjet cartridge as used
   in most consumer inkjet printers. The cartridges usually employ a
   (avoidably) sophisticated design, which often means they are
   protected by patents and therefore cannot be supplied by
   third-party manufacturer at reasonable prices. This makes the
   prolonged use of such printers very cost-ineffective as new
   printers are often sold at prices lower than a new set of
   cartridges would cost.  (Image from: \cite{inkjet_cartridge})}
 \label{fig:inkjet_cartridge}
\end{figure}

While it should be the right of every of every creative mind to be
able to protect their engineering works against plagiarism, securing a
revenue for the ten to twenty years to come, helping to pay back the
expenses and efforts necessary to develop a patented product, it is
very obvious that in the case of ink cartridges, the patents are
solely filed to secure the manufacturers monopoly in the market for
printer supplies. This suspicion is further sustained by the fact,
that companies like HP encourage their customers to send them back
empty inkjet cartridges, claiming that their only intention is to
protect the environment \cite{hp_cartridge_recycling}. However, by
taking care of the recycling of old cartridges, HP eliminates the
viable possibility of refilling and reselling used cartridges by
third-party manufacturers. Thus, HP has gained monopoly in the market
of printer supplies for their products.

Due to their monopoly position, manufacturers like HP, Lexmark and
Canon are able to dictate the prices in the supply market, leading to
absurd prices for their printer ink. The Financial Times reported in
2004 \cite{ft_ink_prices}, that a gallon\footnote{A US gallon
correspondents to $\approx 3.78~\text{liters}$.} of HP's ink for inkjet
printers costs up to 8000 US Dollars, a price that well ranges between
those of synthetic drugs like GHB or artificial insulin \cite{liquid_prices}.

To make things even worse, the overpriced printer supplies have not
only lead to consumer frustration, but also introduced another source
of electronic waste. Since the manufacturers have discovered that they
can much more on supplies than on the actual printers, they have
started subsidizing the printers, allowing them to offer them at
dumping prices as low as 30 Euros, sometimes even undercutting the
price of a new set of cartridges\footnote{Color inkjet printers
usually have to be supplied with at least two inkjet cartridges -
color and black ink - or even four cartridges - 3 colors and black,
each of them costing 10-20 Euros, depending on the type and brand.}
such that many customers actually rather buy a new printer, which
usually ships with a set of ink cartridges, instead of just buying
supplies for the printers that they already own.

Both examples clearly show that proprietary standards can have impact
on not only consumers choices but also have consequences on the
environment, since they prevent the economic (re-)use of consumer
products once they have entered the aftermarket status.

\section{Planned obsolescence}
\label{sec:planned_obsolescence}

Planned obsolescence is a term first coined in the title of Bernhard
London's pamphlet ``Ending the Depression Through Planned
Obsolescence'' in 1932. In his essay, London suggested a regulatory to
be imposed by the government which would subject all consumer products
to a pre-defined, limited lifespan after which the product would have
to be replaced. Upon caught using beyond their expiration date, he
suggested, consumers should even be penalized. His intentions were
motivated by the idea, that a major cause for the \textit{Great
Depression} in the 1930's were consumers' habit of ``using their old
cars, their old tires, their old radios and their old clothing much
longer than statisticians had expected'' \cite{london_obsolescence}.

The principle of planned obsolescence was first put into practice on a
large scale with the formation of the Phoebus cartel in 1924
\cite{phoebus_cartel}. It was a joint-venture of the light bulb
manufacturers Osram, Philips, Tungsram and General Electric, among
others. In the cartel, manufacturers did not only agree on fixing
prices such that no manufacturer had to fear the competition of the
others. But they also secretly signed a commitment, pledging each
other that light bulbs would be designed and constructed such that
their expected life time would not exceed 1000 hours. The Phoebus
cartel demised in 1939 when the war broke out and manufacturers in
Scandinavia decided to start an independent collaboration producing
cheaper light bulbs.

\begin{figure}
\begin{center}
  \fbox{\includegraphics[width=.5\textwidth]{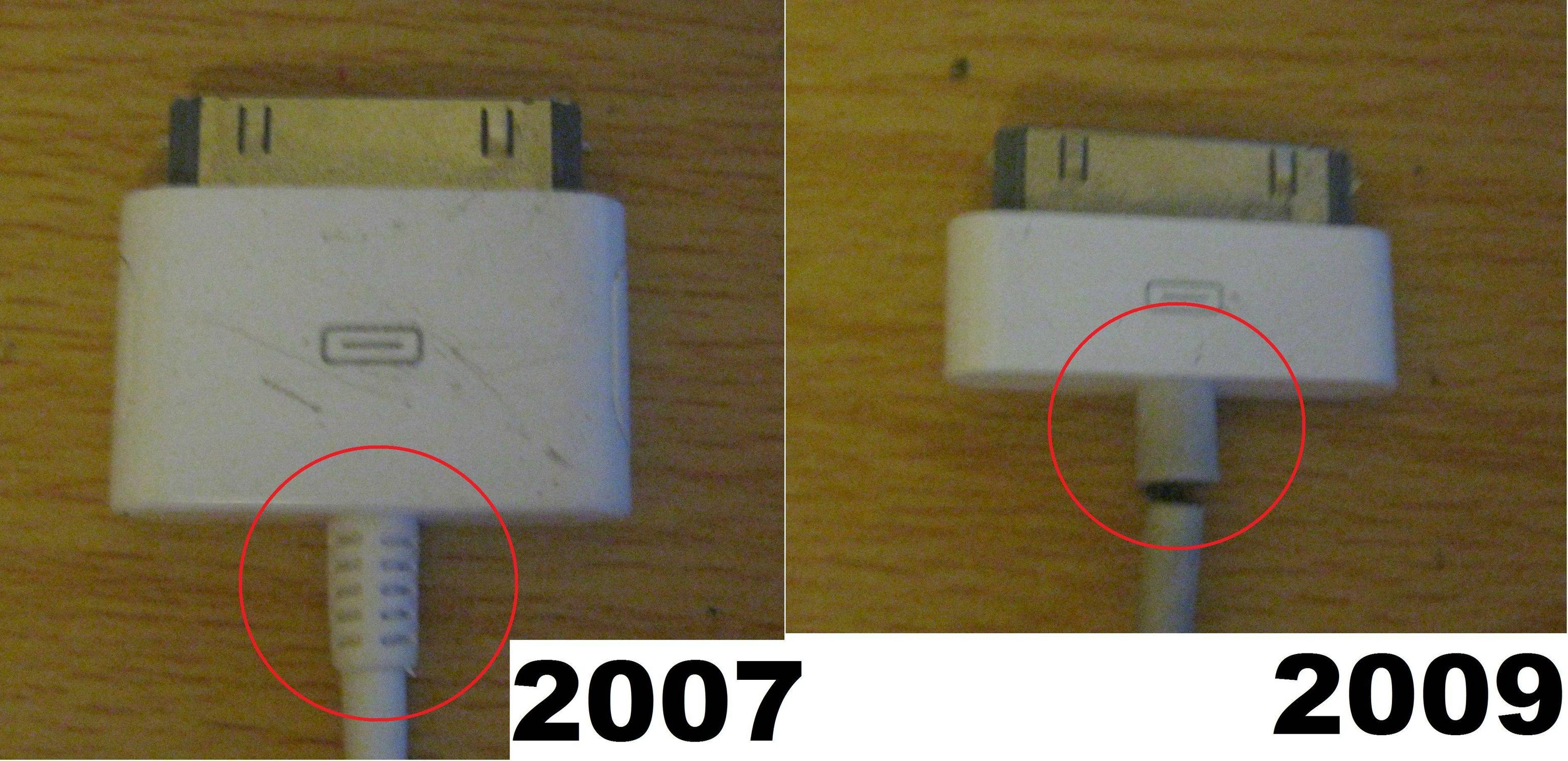}}
\end{center}
 \caption[Consequences of cost reduction]{Consequences of cost
   reduction in the design of electronic consumer products. The image
   shows the standard connection cable of an iPod which has received a
   more cost-effective design over the years, making it more prone to
   failure \cite{reddit_broken_ipod_cable}. However, nowadays many
   consumer products do not obsolete on the hardware level but it's
   rather its software which becomes outdated and thus obsoletes the
   whole product.}
 \label{fig:ipod_broken_cable}
\end{figure}

After its demise, the Phoebus cartel fell into oblivion for a long
time until the controversy it set off became newsworthy with the
advent of modern consumerism: Due to huge advances in design and
manufacturing processes, many consumer products can nowadays be
produced with such high quality standards that they last a life time
and beyond, while still being very cost-effective in
production. The probably most-notable industry worth mentioning in
this context is the semiconductor field. Semiconductors can be
produced very cheap and with very high yield rates, i.e. there are
very little amounts which have to be scrapped. And since a lot of
electronic and even mechanic\footnote{Many electro-mechanic devices
like microphones, switches and relays and much more can nowadays be
produced as semiconductors as well, reducing their cost
dramatically.}functionality can be integrated into a semicondutor,
modern consumer products like cell phones contain nearly only
semiconductors along very few other parts, making them very durable
and cheap. Obsolescence due to technical failure
has therefore become rather uncommon and usually only occur with
design flaws (Fig. \ref{fig:ipod_broken_cable}), which many
manufacturers often recognize leading to free repairs or replacement.

Naturally, the durability of modern consumer products is a problem for
an industry that fuels itself from the sales figures it
generates and companies have to find new ways to constantly motivate
consumers to buy their products. If the consumer products do not
become obsolete by technical failure, they have to obsolete on a
different level. And, in fact, manufacturers have found a viable
method to achieve that and their effective means is to limit the
time the consumer can obtain product support.

While this doesn't really affect conventional consumer products like
household equipment or classical electronic appliances like television
sets and radios, the discontinuation of product support has dramatic
impact for digital products like computers, cell phones (especially
cell phones) and MP3 players. Major parts of the functionality of such
digital products stems from the software they are running. And with
the increase in functionality, the software of digital products has
become more and more complex which makes them more prone to errors in
implementation (often referred to as ``software bugs'') and therefore
subject to frequent software updates \cite{software_update}. Besides
fixing bugs, software updates will very often also bring new
functionality and overall improvements to a digital product.

A very common and prominent example for obsolescence through the lack
of software updates are computers running the popular Microsoft
operating system Windows \cite{microsoft_windows}. Most desktop and
laptop computers ship with the Microsoft operating system already
installed. Since the software has been installed by the hardware
vendor, it is assured that hard- and software work well together and
hardware-specific software, so-called drivers \cite{device_driver},
which are required for the hardware to operate, are working
properly. At the time Microsoft releases a new version of
Windows, the current version will be deprecated and - after a certain
grace time  - fall out of software support by Microsoft because the
old version will have reached the end of its life time
\cite{microsoft_lifecycle}. So, besides the new features which is
users luring into buying the new version of Windows, it is also the
planned obsolescence of the software it which will eventually force
people into the upgrade.

At first sight, one might probably think that planned obsolescence
of software does not eventually lead to large amounts of
electronic waste except for the CDs and DVDs it is stored
on, which are made from polycarbonate, a plastic which
can be easily recycled. However, the opposite is true not in all but
many cases. The problem with an upgrade to a new version of the
Windows operating system is that users will also need to obtain
updated drivers compatible with the new version of Windows. While
updated drivers are usually provided by the hardware vendors for free,
it is not guaranteed that these drivers are provided for older
hardware which is no longer available on the market. The consequence
is that the affected hardware won't work with the new version of
Windows and will have to be replaced with new hardware for which the
hardware vendor provides drivers for the new Windows version. And on
the other hand, for very new hardware drivers aren't usually provided
for older, deprecated versions of Microsoft Windows. So there is
obviously a mutual stimulation between Microsoft and the hardware
drivers which will constantly fuel consumerism which again leads to
the problem of electronic waste.


\section{Consequences: The waste problem}
\label{sec:consequences}

As modern electronics industry encourages end users to replace their
old consumer products with shiny new things from the shelf, the
natural question arises where consumers dispose of their old stuff.

Luckily, there are some options available which allows to give used
electronics a second life by selling it used or donating it for
charity. A very viable option for the first scenario are online market
places like eBay or Craiglist among many others. If someone prefers
donating their used electronics for charity, schools and similar
public facilities are always happy to receive such support. However,
since users of second hardware often run into the same problem of the
lack of software support, many old computers and consumer electronics
end up on waste dumps or incinerators since their recycling is complex
and expensive: It involves a lot of manual work and many steps to
properly disassemble the old equipment such that the component
materials can be recycled or disposed of properly.

\begin{figure}
\begin{center}
  \fbox{\includegraphics[width=.5\textwidth]{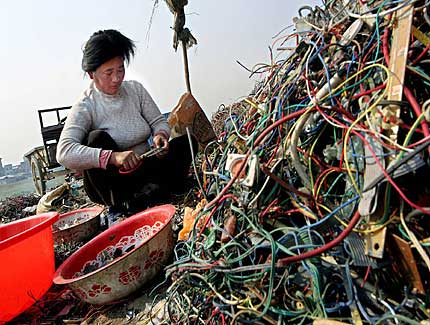}}
\end{center}
 \caption[E-Waste recycling in the Third World]{Electronic waste
   recycling in the Third World: A woman is decomposing electronic
   waste into its raw materials, trying to retrieve valuable
   component materials like gold and copper which they sell to local
   scrap dealers. Due to the recent increase in prices for raw metals,
   more and more people in these countries discover the recycling of
   electronic waste as an additional source of income. Picture from
   \cite{a1gp_ewaste}.}
 \label{fig:ewaste_thirdworld}
\end{figure}

Used electronics contains hazardous materials like mercury, lead,
silver and flame retardants. On the other hand, many parts and
components contain valuable raw materials like gold, copper, titanium
and even platinum, albeit mostly in very small amounts. It can take up
to one ton of electronic waste to recycle just 200 grams of gold
\cite{economicsnp_ewaste}. It is therefore often not viable for
recycling companies in the industrial countries to do the actual
recycling but they found another, much cheaper method for getting rid
of the electronic waste: they export the waste into Third World
countries \cite{greenpeace_ewaste}.

Officially declared as used, working consumer electronics to be sold
on second markets in the Third World, most used electronics turn out to
be non-functioning and useless and therefore end up as electronic
waste. While it is not just immoral to take advantage of these
countries and use them as cheap waste dumps, it also bears hazards for
humans and the environment in the affected countries.

As mentioned before, electronic waste contains many toxic substances
which need to be properly disposed of. This means that decomposing of
the waste has to be performed in appropriately equipped recycling
facilities where the individual component materials of the waste can
be separated from each other in a safe manner such that workers are
never exposed to toxic fumes or liquids. Special waste containments
and protection suits for the workers are essential to guarantee safety
for workers and the environment. Naturally, these requirements are
usually never met and recycling in the Third World usually occurs in
the open, with no protection suits and no respiratory protection,
simply because workers in these countries cannot afford those and are
thus forced to perform the decomposing with their bare hands, often
not knowing about how much these jobs put their health on risk.


\section{Solutions: Stricter laws and consumer awareness}
\label{sec:solutions}

The question now is, what can be done to tackle the waste
problem. Most readers would now demand government action in the form
of stricter laws. Manufacturers should be forced to improve the
recycability of their electronic consumer products, avoiding the use
of toxic materials like lead, arsenic and mercury. Furthermore should
manufacturers be more subject to use standardized connectors and
interfaces, such that, for example, consumers can continue to use the
accessories of their old cell phone once they replace it with a new
one, which avoids rendering the old accessories useless and eventually
electronic waste.

There are, in fact, already certain laws and regulations pending which
enforce such measures. For example, in the European Union, the RoHS
(Restriction of Hazardous Substances) directive \cite{rohs_directive}
has become effective as of July, 1 2006. Often known as the
``lead-free'' directive, it bans the use of lead, mercury, cadmium,
certain types of chromium and some brominated flame retardants. While
lead was primarily used as an ingredient for solder (lead reduces the
brittleness of the tin solder), cadmium has been an essential
ingredient of NiCd rechargeable batteries, which have been superseded
by better and greener battery technologies like NiMH
\cite{nimh_battery} and Li-Ion \cite{lion_battery}. The hazardous
brominated flame retardants have been replaced with non-halogenated
flame retardant, organophosphorous or inorganic flame retardants like
Al(OH)$_3$ or Mg(OH)$_2$.

As for the proprietary connectors, the EU commission has, for example,
requested manufacturers in 2009 to agree on a unified connector for
cell phone chargers \cite{charger_agreement}. This unified connector
is a standard USB connector which can be found on a huge variety of
digital devices. The use of this connector not just allows to use any
cell phone charger independent of the brands of charger and cell
phone, but also give the opportunity to hook up the phone to a normal
PC using a standardized and cheap USB cable which will allow to charge
the phone. The unified connector reduces the amount of chargers users
have to carry around for their various mobile devices and also allows
to continue using a charger even when the cell phone has been replaced
with a new model.

The continued efforts of the European parliament and EU commission to
improve environmental aspects of electronics will certainly bring more
of such directives, agreements and laws, so that, at least for the
European Union, one can say that governments have recognized that
there is a problem with the electronic waste that needs to be taken
care of. Future efforts should certainly concentrate on stricter
export laws for used electronics and encourage manufacturers to be
responsible for the recycling of their used consumer electronics such
that customers can return their devices to the manufacturers once they
have reached the end of their life time.

Since decisions coming from governments side to tackle a problem like
the one with the electronic waste can always take quite a long time
before they become effective after the problem has become imminent
(reasons are usually manufacturers and certain political parties or
countries who block such decisions), it also the responsibility of
consumers to help improving the situation. Consumers can have a very
strong influence on manufacturers environmental guidelines if they
decide to reflect environmental aspects in their decisions of
purchase. In order to help consumers, Greenpeace regularly releases
studies which rank the big manufacturers in electronics industry
according to their environmental efforts
\cite{greenpeace_ranking}. With the help of this ranking and the
included guide, consumers can easily assess manufacturers
environmental efforts and with the decision for greener companies help
making a choice to reduce the amount of toxic electronic waste being
produced. The ranking includes many environmental aspects like the use
of hazardous raw materials, production of climate gases throughout
the manufacturing process as well the recycling efforts shown by the
manufacturers. Clearly, a good starting point for consumers to help.

Also, consumers can prolong the useful life time of their computers
with the help of free and open source software \cite{foss}. For
example, Linux - a free clone of the Unix operating system - supports
a wide range of hardware and especially a lot of old hardware. So
chances are always that an old printer or scanner, for which no device
drivers are available for a current version of Windows, will still
work fine even with the most-recent version of Linux. Popular
distributions of Linux are \textit{Ubuntu} and \textit{Fedora} which
are becoming more and more adopted in private as well as corporate
environments. Together with free desktop software like \textit{Libre
Office} (office suite alternative to Microsoft Office),
\textit{Mozilla Firefox} (a free web browser), \textit{Mozilla
Thunderbird}, \textit{GIMP} (free alternative to Adobe Photoshop)
among many others, consumers not only get a complete set of desktop
applications for free but are also able to use their old computers
with modern software, avoiding the need to replace the hardware once
the operating system is updated.


\section{Conclusion}
\label{sec:conclusion}
Over the last three decades, manufacturers have dramatically improved
the reliability and life time of consumer electronics. However, the
problem of electronic waste is still present since nowadays, very
often (planned) obsolescence does not occur on the hardware level but
on the software level. Software companies and hardware vendors very
often silently stop providing software updates necessary to make old
hardware run with recent software. This is very often the reason why
consumers replace their old electronics with new models, even though
the old electronics is still perfectly functional, just not with
modern software. This problem can be avoided by replacing proprietary
software with free and open source alternatives which very often
provide continued support even for very old hardware. Consumers can
also have a strong voice when coupling their purchase decisions with
environmental aspects. Environment organisations like Greenpeace provide
useful information and guides which will help consumers choose their
consumer electronics while still being able to have a ``green
conscience''.

Governments in Europe have already passed a variety of directives like
the \textit{RoHS} directive and laws which permit the use of hazardous
raw materials, making the recycling of used electronics less harmful
for the environment. Additional agreement with electronics industry
have also been achieved which obligate electronics manufacturers to
use more standardized connectors and parts which help avoid electronic
waste since consumers do not have to throw away all the accessories
when they replace their cell phone, for example. Future efforts of the
governments will have to include a stricter enforcement of export
regulations of used electronics into Third World countries, such that
recycling companies cannot take advantage of these countries as a
cheap source of labor for recycling to harvest valuable raw materials
like gold, copper and other (noble) metals while local workers risk
their health when working without proper protection suits and
respiration masks.

\nocite{*}
\printbibliography
\end{document}